# Design Methodology and Manufacture of a Microinductor

D. Flynn and M. P.Y. Desmulliez
School of Engineering and Physical Sciences, Electrical and Electronic Engineering department, Mountbatten Building, Heriot-Watt University, Edinburgh, UK, EH14 4AS

*Abstract-* Potential core materials to supersede ferrite in the 0.5-10 MHz frequency range are investigated. The performance of electrodeposited nickel-iron, cobalt-iron-copper alloys and the commercial alloy Vitrovac 6025 have been assessed through their inclusion within a custom-made solenoid microinductor. Although the present inductor, at 500 KHz, achieves 77% power efficiency for 24.7W/cm$^3$ power density, an optimized process predicts a power efficiency of 97% for 30.83W/cm$^3$ power density. The principle issues regarding microinductor design and performance are discussed.

I. INTRODUCTION

While the physical size of digital and analogue electronic circuits has been drastically reduced over the past 20 years, the size of their associated power supplies has been reduced at a much slower rate. As a result, the power supply represents an increasing proportion of the size and cost of electronic equipment. One of the main difficulties in the miniaturization of power conversion circuits such as DC-DC converters is the construction of inductors and transformers [1-3].

Recent efforts to miniaturise the overall size of DC-DC power converters, has resulted in an increase in switching frequency of the power circuit from the 100-500 kHz range to the 1-10MHz range. This effort has led to a reduction of the size of the energy storage components that dominate the converter volume. Several problems arise when frequencies are pushed into the MHz region. Core materials commonly used in the 20-500 kHz region such as MnZn ferrites, have rapidly increasing hysteresis and eddy current losses as the frequency is increased. Furthermore, eddy current losses in windings can also become a severe problem, as the skin depth in copper becomes small in relation to the cross section of wire used. Even if these problems are adequately dealt with, the resulting transformer/inductor is still one of the physically largest and most expensive components in the circuit.

Various types of microinductors and transformers using an array of different thin-film metal alloys and winding configurations have been fabricated and reported in the literature over the past 25 years [4-9]. The results of these devices have produced mixed results with varying degrees of success due to the prominent challenge in the design and fabrication of such devices. The key to successful microinductor design is to achieve the optimal balance of the various design parameters for a given application e.g. inductance requirement, low winding resistance, set efficiency at high frequency, high power density etc whilst considering fabrication limitations.

The focus of this paper is to design and fabricate a microinductor that balances the various design tradeoffs yet displays adequate performance for a variety of power related applications. This includes using low-temperature fabrication techniques that permit integration with supporting IC circuitry. The component footprint will be minimized with a high winding packing density limiting the area used while still targeting reasonable inductance values within the 0.5-1 MHz, so that it could be used in a highly miniaturized switched mode power supply for stepping up and down a wide range of voltages. The impact of core geometry on core material properties is highlighted as is the degree of enhancement of the microinductor efficiency and power density following a simple optimization process.

II. INDUCTOR DESIGN

A. Magnetic Core Geometry

A decision on which category of component is optimal requires a comparison across a variety of factors such as power handling, efficiency, practicality of fabrication, inductance requirement etc. Consider a core material for a one-turn micro-inductor. The saturation flux density, $B_{sat}$,

$$B_{sat} = \frac{\mu_o \mu_r NI}{l_c} \qquad (1)$$

is a predetermined constant for a given material. In the same way, the current applied to the component would be known for a given application. Therefore, the only free parameter left will be either the relative permeability or the flux path length, since

$$\frac{l_c}{\mu_r} = \frac{I\mu_o}{B} \qquad (2)$$

The core, typically a rectangle, triangle or a circle, has a certain perimeter. Considering a rectangular core within a one turn pot-core component, shown in Fig. 1 with the main flux path indicated. As frequency increases the skin depths of the magnetic material and winding decrease. Therefore, to minimise skin depth losses and maintain the inductance value i.e. overall magnetic core area, the device becomes thinner and wider. The resulting increase in flux path length requires an increase in relative permeability for the inductance value to be maintained. An increase in relative permeability for the same resistivity value will make the material susceptible to skin depth losses at reduced frequency.

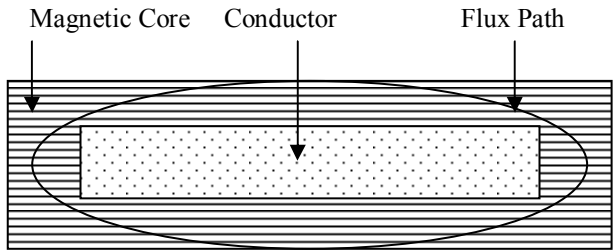

Fig. 1. Simple rectangular one-turn micro-inductor.

If, for the same area and core material, the rectangular geometry was changed to a triangular geometry, then the perimeter is reduced by one third. The required permeability is then reduced by 1/3 and greater skin depth is achieved for the same resistivity.

A one turn micro-inductor with a triangular trench structure was fabricated by Sullivan et al [10]. The fabrication of this V-groove micro-inductor required sputter deposition. Sputtering is a slow and costly process, and due to delamination can only be used for small core areas. Section will demonstrate how limiting the number of laminate layers can have an adverse effect on component performance.

A circular component would suffer from the same technical disadvantages as the triangular component. Moreover, a circular component fabricated by micromachining methods would require non-planar construction. Minimal footprint and low profile are prerequisites for present and future magnetic components. Therefore, a circular component was not investigated.

Due to the aforementioned complications of triangular and circular geometries. the investigated components herein are of rectangular planar form. Disadvantages of this geometry will be overcome via tailored core properties and optimal component design.

Another consideration is the tailoring of magnetic film properties with magnetic field annealing. Both pot-core and solenoid geometries can be used with magnetic field annealing; therefore, this is not a design restriction.

B. Winding Geometry

The winding geometries generally encountered for the inductor are of the spiral, meander or solenoid forms. A single layer solenoid component allows near ideal performance [11]. The number of turns per single layer of traditionally wound solenoid inductors is limited by the core's inner diameter and how tightly you can pack the turns together. To increase inductance, a larger core or multiple layers of windings would normally be needed. This would normally lead to an increase in component size and parasitic capacitance. UV-photolithography discussed within the manufacturing of the solenoid allows a higher winding density avoiding such methods.

Sometimes an important issue in choosing a winding configuration is the magnitude and distribution of external magnetic fields generated by the transformer or inductor. The conventional pot-core is regarded as performing more favorably in minimizing external magnetic fields, as the core encloses and so shields the windings. The result is low external fields regardless of the distribution of primary and secondary windings inside the core. Solenoid components also have low external fields due to the winding distribution. In transformer applications the primary and secondary windings are customarily distributed evenly around the core, lying on top of each other with little space between. Thus the currents cancel each other, so little external field results. In micro-fabrication the ability to finely pattern windings makes it relatively easy to arrange the windings so that primary and secondary currents locally cancel. Hence, external fields are not a primary concern in component development within this work.

Perhaps the most important factor in comparing the pot-core, spiral type winding, and solenoid designs is the ease of manufacture. This parameter is the most difficult to quantify and obviously dependent on fabrication capability. In the solenoid case, a difficulty often encountered is the interconnection of the top and bottom winding layers to encircle the core. This requires accurate alignment between the layers to insure proper connections, requires low-resistance contacts and connection over a vertical distance greater than the thickness of the core. A novel flip-chip bonding procedure overcomes the aforementioned difficulties [12].

In the pot core design, the inter-layer connection requirements are less severe because the windings are planar and manufactured in a single deposition process. A complication arises if the inductor winding is spiral as access to the output pad is required. This typically requires a fourth deposition step, or the use of external connections such as bond wires. The thickness of the conductor layers is also typically larger than the thickness of the core layers, therefore, the vertical distance over which connections must be made is larger in the pot-core design. Furthermore, to form a closed magnetic core the upper and lower sections of the core have to be joined. The points where the two layers join is within the main flux path and can generate an unfavorable leakage flux leading to an increase in winding resistance, electromagnetic interference (EMI) and a reduction in component inductance.

C. Analytical Design FlowChart

Fig. 2 displays the design flowchart for the development of a microinductor. The solenoid microinductor developed within this work formed a closed magnetic path. The accuracy of the analytical equations will be reduced with components that include air-gaps within the main flux path due to fringing effects. The equations within Fig.2 are used for the theoretical data within section III.

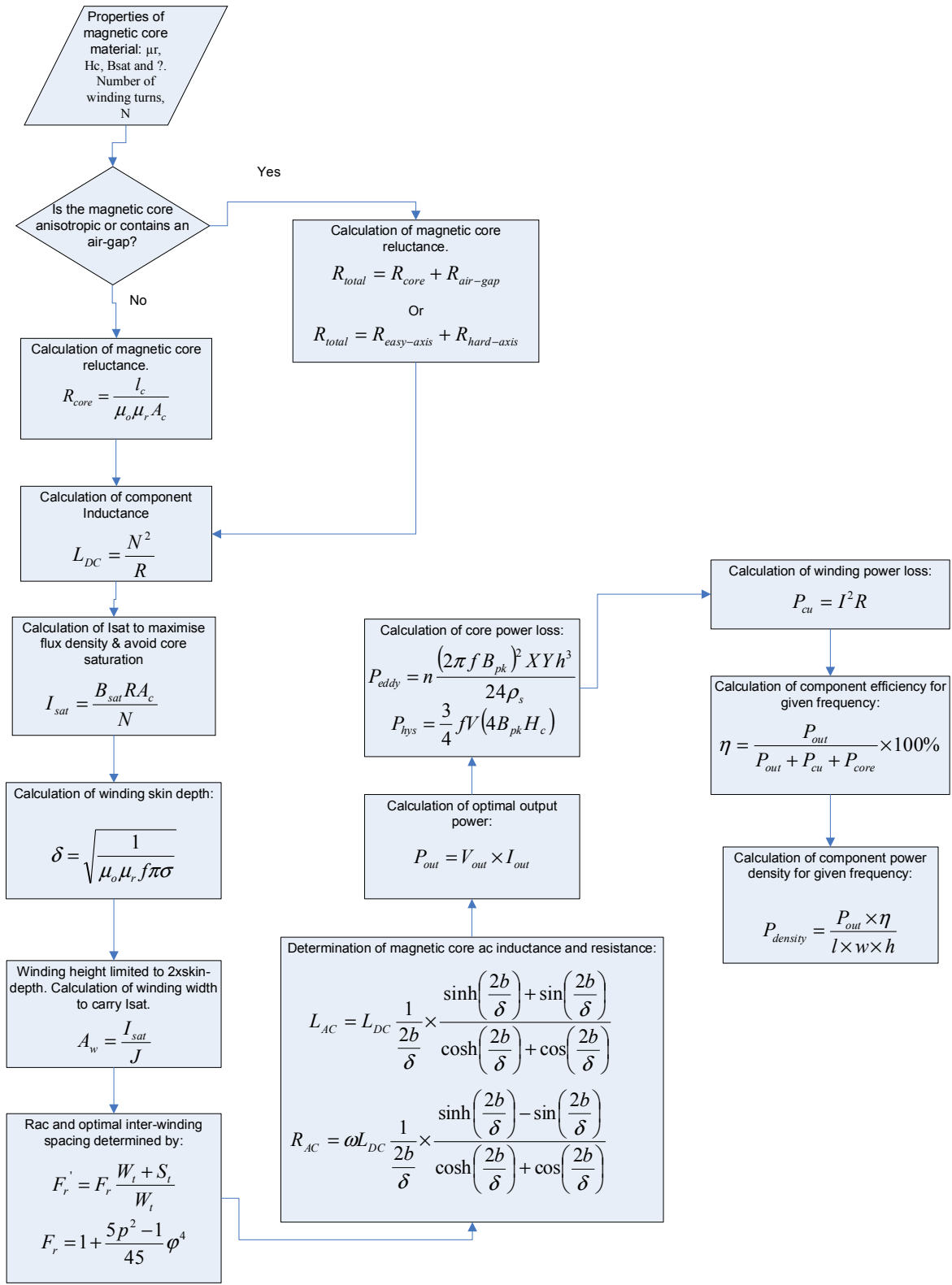

Fig. 2 The Design Flowchart of the Microinductor

III. COMPONENT CHARACTERISATION

The novel microinductor fabricated via flip chip assembly is shown in Fig.3. The manufacture process of the inductor and properties of the core materials utilized are described in detail within [13]. The (*) symbol is used to denote magnetically anisotropic films, as displayed in Fig.3.

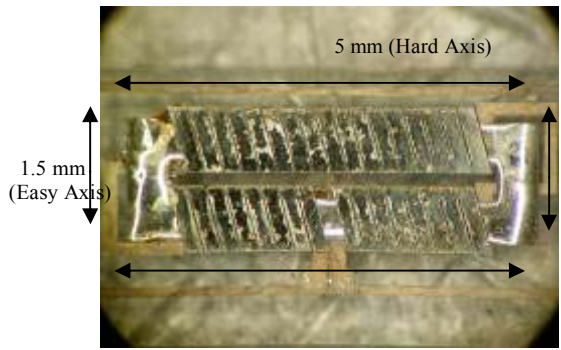

Fig. 3. Fabricated micro-inductor. The O-shape core is assembled between the windings layers prior to flip-chip bonding. The component is approximately 5mmx2mmx0.25mm (*lxbxh*), respectively. When Ni-Fe is anisotropic the orientation of the Easy and Hard axes are indicated.

A critical point in the performance of an inductor is the measurement of the inductance value as a function of frequency and the additional core losses. A Hewlett Packard 4192A LF-impedance analyzer was used to record the inductance, resistance and Q-factor over a frequency range of 1 kHz-10MHz. The resistance value shown in Fig.4 is for an air-core component. The reason for omitting the magnetic core was to remove core AC resistance from the recorded value. Therefore, the increase in resistance at high frequency from the $R_{DC}$ value is due to skin and proximity affects with increasing frequency. The winding is 90µm thick and at 1MHz has a skin depth value of approximately 66µm. Due to the winding consisting of a single layer and inter-winding spacing optimized, the proximity effect is minimized and the main contributor to the ac resistance is assumed to be the skin effect.

One important source of loss in a dc-dc converter is the dc winding resistance, $R_{dc}$, of the output inductor. This parameter defines the minimal loss condition of the inductor. For example, if $R_{dc}$ = 1mΩ, and dc output current of *I*=100 Amps the inductor dc power loss is;

$$P = I^2 R_{dc} = 10(W)$$

If the converter were to provide 100 Amps at 0.75 Volts, the loss in efficiency due to the inductor $R_{dc}$ would be 11.76 %. This result demonstrates that the converter would be less than 89% efficient due to $R_{dc}$ alone. Hence, it is important to have a winding construction that minimizes $R_{dc}$ whilst still meeting applied current and inductance criteria.

The resistance of the components with the magnetic core included are shown in Fig.5. The minor variation between experimental and analytical values may be due to the limitations of the 2D model representing the intrinsically 3D nature of the leakage and eddy current flux.

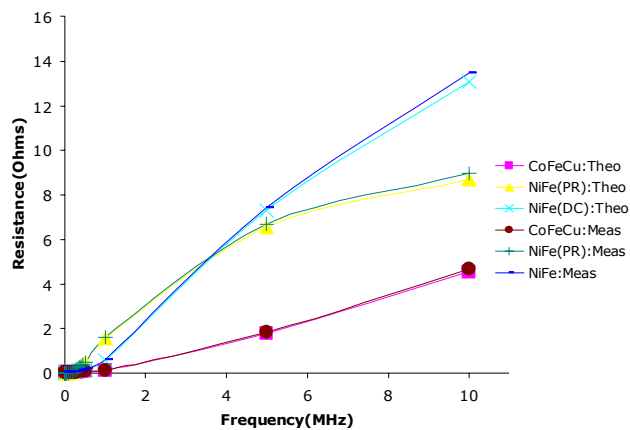

Fig. 5. Micro-inductor resistances as a function of frequency. The resistance represents the winding and core contribution.

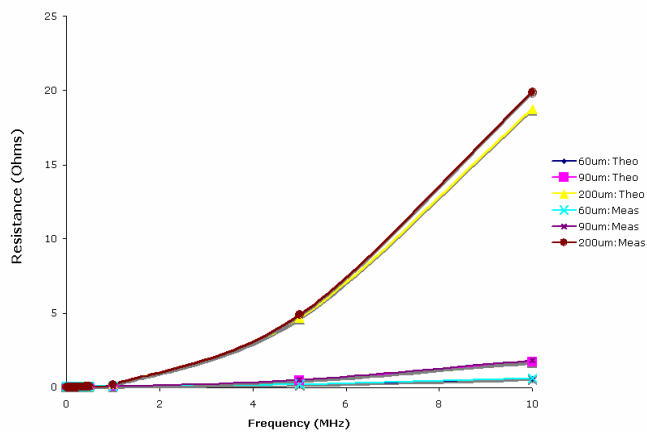

Fig.4. Winding resistance vs. frequency for 90µm thick copper windings

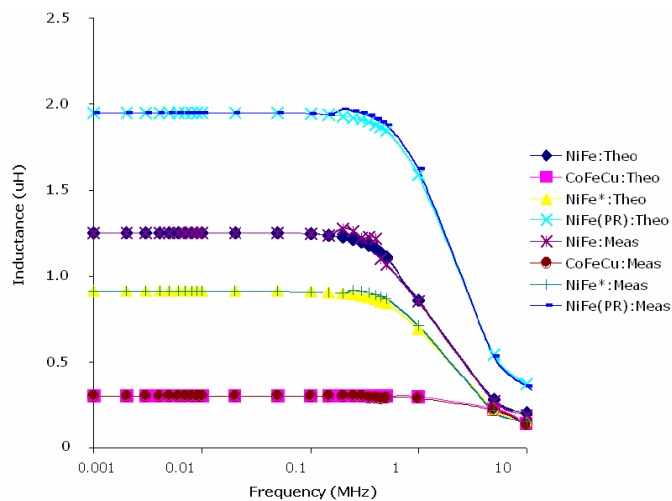

Fig. 6. Inductance vs. Frequency of the electrodeposited alloys

Fig.6 displays the theoretical values (theo) calculated and measured values (meas) of inductance with increasing frequency.

The self resonant frequency (SRF) of an inductor is defined as the frequency at which the reactance becomes zero, (null Q-factor). The measured reactance increases linearly with frequency; hence the SRF induced by parasitic capacitance effects does not occur in the frequency range of interest. The Q-factor of the electrodeposited alloys is displayed in Fig.7. The maximum Q-factors of the alloys occurred in the 500 kHz-1MHz frequency range indicating that the thin film dimensions are acceptable for operation within this range. Due to the properties of the CoFeCu core the 10μm thick core layer remains below one skin depth at 1MHz. The thicknesses of the components windings are under 2×skin depth therefore AC winding losses are maintained at acceptable levels. Therefore the ratio of the stored power to dissipative power of the component, the Q-factor, was optimal in this frequency range.

The Q-factor response of the commercial alloy demonstrates the susceptibility of this particular film to skin depth effects. The skin depth effects increase the dissipative losses within the component, therefore reduce the Q-factor. The high Q-factor in the low frequency range is greatly reduced at 1 MHz with respect to the CoFeCu film.

Using the theoretical DC saturation current formula within Fig.2. the maximum applied current can be determined. The performance of the prototypes is summarized in Table 1. K is taken to be equal to 4.44, sinusoidal waveform, and $B_{sat}$ is used to maximize the power density of the component.

TABLE 1
MICROINDUCTOR PERFORMANCE FOR RESPECTIVE CORES AT 0.5MHZ

| 500kHz Parameter | NiFe | NiFe* | Vit | NiFe (PR) | CoFeCu |
|---|---|---|---|---|---|
| Isat(mA) | 100 | 137 | 1.32 | 66 | 180 |
| $L(\mu H)$ | 1.07 | 0.86 | 15.5 | 1.87 | 0.28 |
| $V_{IN}$ | 0.29 | 0.29 | 0.36 | 0.28 | 0.51 |
| Pout (mW) | 29 | 40 | 0.47 | 18.48 | 91.8 |
| Peddy (mW) | 7.23 | 7.23 | 3.47 | 2.86 | 24.6 |
| Phys (mW) | 0.198 | 0.03 | 0.02 | 0.08 | 0.11 |
| Pcu (mW) | 0.5 | 0.9 | 0.00008 | 0.21 | 1.62 |
| Efficiency (%) | 78 | 82 | X | 85 | 77 |
| Power Density (W/cm³) | 7.8 | 11.3 | X | 5.46 | 24.7 |

The saturation current of the Vitrovac material is only 1.33mA. Hence, high efficiency is achieved at the expense of power density. An appropriate reduction in the number of windings and increase in core area via lamination will result in the same overall inductance. This would reduce the major Joule loss mechanism, winding loss, and allow windings of larger dimensions to be fabricated in order to apply greater current. The 90μm thick windings limit the maximum current to 180mA which is well beneath the saturation current of the CoFeCu components. An optimal component design process is therefore required.

IV. OPTIMAL DESIGN PROCESS

A simplified optimal design procedure for the solenoid component is outlined within this section. The optimization is performed in two circumstances in reference to the CoFeCu prototype in section III; (1) the aim is to maintain the inductance value whilst increasing the power density significantly and improve efficiency, and (2) the primary aims are two maintain inductance and improve efficiency. With good relative permeability, small coercivity, reasonable resistivity and the largest saturation current (1.24A), the CoFeCu alloy is selected for optimization. Once a frequency is selected, in this case 0.5 MHz, skin depth values for the windings and core laminate are 93μm and 17.4μm, respectively. For comparison with the prototype, the inductance required will be 0.3μH (Fig.12). The core area and numbers of turns are therefore:

$$LI = NA_cB \qquad (3)$$

Therefore the area of the component is:

$$A_c = \frac{LI}{BN} \qquad (4)$$

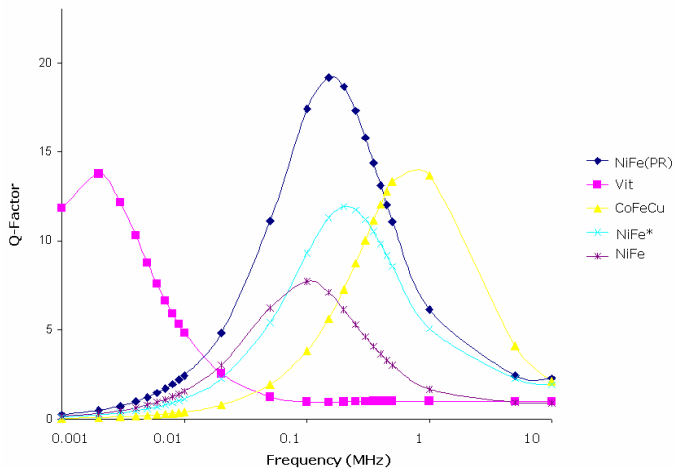

Fig. 7. Q-factor vs. frequency for the electrodeposited alloys

Design constraints are included at this stage. Each CoFeCu laminate will equal ½ skin depth and the number of laminations will be restricted to 10. The number of laminations will affect component manufacturability and cost such that N is only 11 turns in our case. The maximum applied current is 1.24A, PCB current density is 10A/mm² and winding height is restricted to 2xδ. The dimensions and performance of the optimized component can therefore be calculated and are given in Table 2.

TABLE 2
OPTIMISED PERFORMANCE & PARAMETERS

| Dimensions | Optimized component (1) | Optimized component (2) | Prototype Component |
|---|---|---|---|
| Number of turns | 11 | 33 | 33 |
| Turn thickness | 180 µm | 90 µm | 90 µm |
| Turn width | 550 µm | 200 µm | 200 µm |
| Turn spacing | 55 µm | 20 µm | 20 µm |
| Laminations | 2 | 10 | 1 |
| Lamination thickness | 5 µm | 1 µm | 10 µm |
| Lamination width | 500 µm | 500 µm | 500 µm |
| Lamination insulation | 5 µm | 5 µm | 5 µm |
| Performance | Optimized component | Prototype Component | Prototype Component |
| $V_{in}$ | 0.17 | 0.51 | 0.51 |
| $P_{out}$ | 0.2108 (W) | 0.0918 (W) | 0.0918 (W) |
| $B_{pk}$ | 1.4(T) | 1.4(T) | 1.4(T) |
| $P_{eddy}$ | 5.54(mW) | 0.22(mW) | 24.6(mW) |
| $P_{hys}$ | 0.57(mW) | 0.11(mW) | 0.11(mW) |
| $P_{Cu}$ | 184(mW) | 1.62(mW) | 1.62(mW) |
| Efficiency (%) | **83** | **97** | **77** |
| Power density (W/cm³) | **174.8** | **30.83** | **24.7** |

Optimized component (1) produces a 6% increase in efficiency and 7 fold increase in power density. Conventional inductors and transformers will normally operate with efficiency in the range of 90-95%. Hence, the application would have to greatly benefit from the increase in power density to tolerate the level of efficiency. Optimized component (2) uses the data from the prototype to identify eddy current core loss as the main loss mechanism. The core is reduced to ten 1µm laminates and via this adjustment improves efficiency by 20%. Comparing the two optimized components, a trade off between efficiency and power density is evident.

## V. CONCLUSION

Micromachined inductors with different magnetic cores have been fabricated on glass using micromachined techniques borrowed from the LIGA process. The analytical results using the two dimensional model agree well with the experimental values of inductance.

The challenge of developing inductors and transformers for MHz DC-DC converter operation requires the development of suitable core materials and manufacturability of laminated layers. This fact is demonstrated by the results within Tables 1 and 2. The micro-inductor presented in this article has proved that an efficient component can be manufactured via an inexpensive UV LIGA and electroplating process.

ACKNOWLEDGMENT

Thanks are extended to Prof Cywinski for the VSM data performed at the University of Leeds, and to Raytheon Systems Limited for performing the resistivity measurements. This work was made possible through the funding of the Scottish Consortium in Integrated Micro Photonic Systems (SCIMPS) funded by the Scottish Funding Council under the Strategic Research Development Grant scheme.